\documentclass[12pt]{article}
\def\gappeq{\mathrel{\rlap {\raise.5ex\hbox{$>$}}
{\lower.5ex\hbox{$\sim$}}}}

\def\lappeq{\mathrel{\rlap{\raise.5ex\hbox{$<$}}
{\lower.5ex\hbox{$\sim$}}}}

\def\Toprel#1\over#2{\mathrel{\mathop{#2}\limits^{#1}}}

\begin{document}
\pagestyle{empty}
\begin{flushright}
{CERN-TH/2001-030}\\
hep-ph/0102223
\end{flushright}
\vspace*{5mm}
\begin{center}
{\Large  \bf Light-quark masses and renormalons}\\
\vspace*{1cm} 
{\bf A.L. Kataev}\\
\vspace{0.3cm}
Theoretical Physics Division, CERN CH-1211, Geneva 23, Switzerland  \\
and Institute for Nuclear Research of the Academy of Sciences of Russia,\\
117312 Moscow, Russia
\end{center}
\vspace*{2cm}  
\begin{center}
{\bf ABSTRACT} 
\end{center}
\vspace*{5mm}
\noindent
A brief  review  of the problem of the  determination  of light-quark 
masses from QCD sum rules for the correlators of scalar 
and pseudoscalar currents is given. Special attention is paid 
to the description of the 
large-$N_f$ results, obtained in  collaboration with Broadhurst and Maxwell, 
for the scalar correlator, and estimates of its higher-order 
perturbative uncertainties within the 
renormalon-inspired large-$\beta_0$ expansion in the 
$\overline{\rm MS}$ scheme are given. Brief discussion is 
presented  of the results 
of calculations of higher-order perturbative QCD corrections to 
the relation between pole and   
$\overline{\rm MS}$-scheme running-quark 
masses. Their comparison with the different estimates is also given.
\vspace*{0.5cm} 
\noindent 
%\rule[.1in]{16.5cm}{.002in}

\begin{center}
{\it Contributed to the Proceedings of Quarks-2000 
International Seminar,
\\ May 2000, Pushkin,  Russia}
\end{center}
\vspace{0.5cm}

\begin{flushleft} CERN-TH/2001-030 \\
February 2001
\end{flushleft}
\vfill\eject
%\pagestyle{empty}
%\clearpage\mbox{}\clearpage

\setcounter{page}{1}
\pagestyle{plain}

%\documentclass[12pt]{article}
%\begin{document}

%\documentclass[12pt]{article}
% Or, if you are still using LaTeX2.09
%\documentstyle[12pt]{article}
%\begin{document}
\begin{center}
{\Large  \bf Light-quark masses and renormalons}

{\bf A.L. Kataev}

{Theoretical Physics Division, CERN CH-1211, Geneva, Switzerland \\
and Institute for Nuclear Research of the Academy of Sciences of Russia,\\
117312 Moscow, Russia}

\end{center}

\begin{abstract}
A brief  review  of the problem of the  determination  of light-quark 
masses from QCD sum rules for the correlators of scalar 
and pseudoscalar currents is given. Special attention is paid 
to the description of the 
large-$N_f$ results, obtained in  collaboration with Broadhurst and Maxwell, 
for the scalar correlator, and estimates of its higher-order 
perturbative uncertainties within the 
renormalon-inspired large-$\beta_0$ expansion in the 
$\overline{\rm MS}$ scheme are given. Brief discussion is 
presented  of the results 
of calculations of higher-order perturbative QCD corrections to 
the relation between pole and   
$\overline{\rm MS}$-scheme running-quark 
masses. Their comparison with the different estimates is also given.
\end{abstract}

\section{Introduction}
The proposal to  extract the values of  light-quark current  masses 
from the QCD sum rules for the correlators 
of quark scalar and pseudoscalar currents  is rather old (see e.g. 
the works of 
Refs.~\cite{Vainshtein:1978nn}--\cite{Dominguez:1987aa}
and the reviews of Refs.~\cite{Gasser:1982ap}--\cite{Narison:1987da}).
However, in view of the appearance of new experimental data and new 
theoretical results and ideas, 
the investigations in this direction are continuing
(for a summary  of the modern situation, see forinstance the 
reviews of  Refs.~\cite{Narison:2000uj,Gupta:2001cu}). 

When studies began, it was not clear what normalization point should be 
associated with the definite values  
of the  light-quark current masses. Starting  from the  mid-90,s 
it became common wisdom, accepted by the Particle Data Group, to 
consider as phenomenolgical parameters the running quark 
masses of the $\overline{\rm MS}$ scheme, normalized at  either 
1~GeV or 2~GeV. Another quantity, probably more suitable for the calculations 
on the lattice, is the renormalization-scheme-invariant mass $\hat{m}$, 
connected with the running one by the following 
solution of the renormalization group 
equation 
\begin{eqnarray} 
\overline{m}(Q^2)&=& \hat{m}~ \rm exp\bigg[-\int^{\alpha_s(Q^2)}_0
\frac{\gamma_m(x)}{\beta(x)}dx-\int^{2\beta_0}_0\frac{\gamma_0}{\beta_0 x}dx
+2\frac{\gamma_0}{\beta_0}\rm{ln}(2\beta_0)\bigg]
\\ \nonumber 
&=& \hat{m}(2\beta_0\alpha_s(Q^2))^{\gamma_0/\beta_0}[1+....]~~,
\end{eqnarray}
where the perturbative QCD $\beta$-function and the mass-anomalous dimension 
function $\gamma_m$ 
are now known at the 4-loop approximations (see Ref.~\cite{vanRitbergen:1997va}
and Refs.~\cite{Chetyrkin:1997dh, Vermaseren:1997fq} respectively) with the 
one-loop scheme-independent terms $\beta_0=[11-(2/3)N_f]/4$ and $\gamma_0=1$.

Another important definition of the current masses is  the 
on-shell or pole quark masses. Despite the fact that, contrary 
to the case of charged leptons, quarks do not exist as free particles, 
this definition is commonly accepted and widely used when heavy quarks 
are considered.  
Within perturbation theory and the   
$\overline{\rm MS}$ scheme, the running quark masses and 
pole quark masses can be straightforwardly connected. 
At the next-to-leading 
order (NLO) this relation was calculated in Ref.~\cite{Gray:1990yh} and confirmed later on in Ref.~\cite{Fleischer:1999dw}.
Beyond the NLO level, higher order terms to this relation 
were estimated 
in the large-$\beta_0$ approximation 
\cite{Beneke:1995qe,Philippides:1995jw}, 
which can be obtained 
from the gauge-independent large-$N_f$ expansion  after 
renormalon-inspired substitution $N_f\rightarrow -6\beta_0$.
At the NNLO level the correction 
to the relation between pole and running 
quark masses 
was estimated in Ref.~\cite{Chetyrkin:1997wm} within the 
framework of the scheme-invariant approach developed in the works 
of Ref.~\cite{Kataev:1995vh}. 
 
The results of the subsequent  approximate NNLO calculations, 
made by evaluating 
expansions of the  3-loop quark self-energy for small and large 
external momentum and further 
Pad\'e improvements of the corresponding mass-dependent series
\cite{Chetyrkin:1999ys}, turned out to be in reasonable agreement 
with estimates, obtained by both large-$\beta_0$ resummation 
\cite{Beneke:1995qe,Philippides:1995jw} and scheme-invariant 
methods \cite{Chetyrkin:1997wm}. In its turn, the numerical result
of the approximate calculation of Ref.~\cite{Chetyrkin:1999ys}
was confirmed by the recent explicit analytical calculations, 
performed in Ref.~\cite{Melnikov:2000qh}.

In this report the  survey of the uncertainties   
of recent QCD sum-rule determinations 
of light-quark masses from the correlators 
of scalar and pseudoscalar currents at 3-loop  \cite{Bijnens:1995ci}--  
\cite{Jamin:1995vr}
and 4-loop \cite{Chetyrkin:1997xa} perturbative QCD approximations
is given. Indeed, 
a single-chain renormalon structure 
of the Borel images of the functions, related to 
the correlator of scalar quark currents was recently studied in collaboration 
with Broadhurst and Maxwell in  
Ref.~\cite{Broadhurst:2001yc}. In this note  special attention is paid   
to the  results obtained in Ref.~\cite{Broadhurst:2001yc}. The higher-order 
perturbative QCD uncertainties for the studied  functions are estimated with 
the help of the large-$\beta_0$ approximation. 
The bridge between the analysis of  similar questions 
in the case of the relation between pole and running quark masses, 
which contain the renormalon singularity \cite{Bigi:1994em,Beneke:1994sw},
will also be  constructed.
The reason for  the limitations  of the 
scheme-invariant estimates of Ref.~\cite{Chetyrkin:1997wm} in the scalar 
channel 
will be clarified. 

\section{Renormalon calculus}

The studies of renormalon effects (for  
reviews see, \cite{Altarelli,Beneke:2000kc})  usually  start from the 
consideration of the contributions  of the internal single chain of fermion
loops  
to some physical quantity $D$, say to the  anomalous magnetic moment 
of the electron \cite{Lautrup:1977hs}
or 
to  the $e^+e^-$-annihilation 
Adler function 
\cite{Zakharov:1992bx,Brown:1992pk}. 
From a perturbative point 
of view, these insertions form the basis of the  
large-$N_f$ expansions, which are  widely used at present. 
In principle, the coefficients 
of these  expansions  are scheme-dependent 
(for the discussion of the scheme dependence of renormalon 
contributions, see Ref.~\cite{Krasnikov:1996jq}), but in what follows 
this problem will not  be  analysed since   
the $\overline{\rm MS}$ scheme will be essentially used. 

The next step in the renormalon calculus is the construction,   from a 
perturbative expansion of the  physical quantity $D(a)$
\begin{equation}
D(a)=\sum_{n\geq 0}d_n a^{n+1}
\end{equation}
of the Borel transform 
\begin{equation}
B[D](\delta)=\sum_{n\geq 0}d_n\frac{\delta^n}{n!}~~,
\label{BT}
\end{equation}
and of   the Borel integral 
\begin{equation}
\tilde{D}(a)=\int_0^{\infty}e^{-\delta/a} B[D](\delta)~d\delta~~.
\label{BI}
\end{equation}
The basic step  in defining this integral is related to the 
analysis of the singularities of the Borel transform $B[D(\delta)]$.
In the case when  it has no singularities on the  real positive axis $\delta$,
the integral of Eq.~(\ref{BT}) is well defined.
The singularities, located  on the negative 
axis $\delta$, are called {\it ultraviolet} (UV) {\it renormalons}; after 
Borel resummation, they generate the sign-alternating perturbative 
QCD 
series with 
factorially increasing coefficients. The singularities, located 
on the positive axis $\delta$ are called {\it infrared } (IR) 
{\it renormalons}. 
These  are associated with the sign constant QCD  perturbative  
expansions with coefficients, increasing like $n!$. 
In this case the Borel integral is ill-defined, 
but its Cauchy Principle Value (PV) can be used (for one of the first 
applications to the resummation of the renormalon-chain, 
perturbative contributions 
to  the $(g-2)/2$ of electron, in QED  \cite{Lautrup:1977hs} 
with coefficients growing like $(+1)^nn!$, 
see Ref.~\cite{Ilchev:1981mh}).
In fact, 
it is possible to show that 
the IR renormalon singularities  can be associated to power-suppressed 
corrections of order $(\Lambda/Q)^{(2p)}$, where $p$ is the number, related 
to  the 
pole singularity, at the points $\delta$ of 
the  positive axis of the Borel plane (see e.g. Ref.~\cite{Beneke:2000kc}).

Up to now all discussions were made within the framework of large-$N_f$ 
expansion, which is well-defined in the cases of both QED 
and QCD. It should be stressed that,  since the first coefficient 
of the QED $\beta$-function is proportional to $N_f$, the large-$N_f$ limit 
of QED is identical to the large-$\beta_0$ approximation. In the case 
of QCD the situation is more ambiguous. Indeed, in order  
the rewrite the large-$N_f$ perturbative series in the form 
of the large-$\beta_0$ approximation  the {\it naive 
nonabelianization} ansatz \cite{Broadhurst:1995se}, namely 
the substitution $N_f\rightarrow N_f-11N_c/2$=$-6\beta_0$
should be used.  This  procedure 
turned out to be rather useful in the analysis of both renormalon 
effects and estimates of the uncertainties due to perturbative 
corrections beyond 
the calculated ones (see e.g. Ref. \cite{Lovett-Turner:1995ti}).

\section{Definitions of the basic quantities}    

Consider now the correlator of scalar quark currents 
$J_S=m(\mu)\overline{\psi}\psi$, namely 
the Green function 
\begin{equation}
\Pi_S(Q^2)=i\int e^{iqx}\langle J_S(x) J_S(0)\rangle_0 d^4x~~~.
\end{equation}
Its imaginary part ${\rm Im}\Pi_S=2\pi s R_S\left(1+O(1/s)\right)$ is related to the spectral 
function $R_S(s)$ of two Euclidean constructs, considered 
in the work of Ref.~\cite{Broadhurst:2001yc}:                                  
\begin{equation}
\tilde{D}_S(Q^2)=Q^2\int_0^{\infty}\frac{R_S(s)~ds}{(s+Q^2)^2}
\end{equation}
\begin{equation}
\overline{D}_S(Q^2)=2Q^2\int_0^{\infty}\frac{s R_S(s) ds}{(s+Q^2)^3}~~~.
\end{equation}
The first one is defined through the  dispersion relation for the 
first derivative of $\Pi_S(Q^2)/Q^2$, while the second one is defined 
by the {\it  second} derivative for $\Pi_S(Q^2)$.
The $\tilde{D}(Q^2)$-function
was used in the process of calculating
the 2-loop and 3-loop QCD corrections to $R_S(s)$ in 
the $\overline{\rm MS}$ scheme in Refs.
~\cite{Gorishnii:1984cu}-\cite{Gorishnii:1991zr}. 
The 4-loop QCD logarithmic corrections to $\Pi_S$, and therefore  the 
$\alpha_s^3$ contributions to its imaginary part 
$R_S(s)$, were evaluated  in Ref.~\cite{Chetyrkin:1997sr}.
In the process of these calculations the following inhomogeneous 
renormalization group equation was used:
\begin{equation}
\bigg(\frac{\partial}{\partial\rm{log}\mu^2}+\beta(\alpha_s)
\frac{\partial}{\partial\alpha_s}+2\gamma_m(\alpha_s)\bigg)\Pi_S=
[m(\mu^2)]^2 Q^2 \bigg[ \gamma_{SS}(\alpha_s)+O\left(\frac{m^2}{Q^2}\right) \bigg]~~~.
\end{equation}
In leading order in $1/Q^2$, and next-to-leading order in $1/N_f$,
we have \cite{Broadhurst:2001yc}:
\begin{eqnarray}
\Pi_S&=&[m(\mu^2)]^2Q^2d_F\bigg(-2L-4+\frac{C_Fb}{T_fN_f}H(L,b)+
O\bigg(\frac{1}{N_f^2}\bigg)\bigg) \\ \nonumber 
\gamma_{SS}&=&d_F\bigg(-2+\frac{C_F b}{T_fN_f}h(b)+O\bigg(\frac{1}{N_f^2}
\bigg)\bigg)~~,
\end{eqnarray}
where $d_F=3$, $C_F=4/3$, $L=\rm ln (\mu^2/Q^2)$ and $b=T_fN_f\alpha_s/(3\pi)$.
The expression of the coefficients $H_n(L)$ of the large-$N_f$ expansion 
of $H(L,b)$, namely  $H(L,b)$=$\sum_{n> 1} H_n(L)b^{n-2}$, can be obtained 
from the all-order solution of the following relation~\cite{Broadhurst:2001yc}:
\begin{eqnarray}
n(n-1)H_n(L)&=& n\bigg(h_{n+1}-4(L+2)g_n\bigg)+4g_{n+1}-9(-1)^nD_n(L) \\ 
\nonumber
\sum_{n}\frac{D_n(L)\delta^n}{n!}&=&\bigg[1+\delta G_D(\delta)\bigg]
\rm{exp}\bigg(\bigg(L+\frac{5}{3}\bigg)\delta\bigg)~~~,
\end{eqnarray}
where $h_{n}$ are defined by the large-$N_f$ expansion of the  $h(b)$-function 
in the expression for $\gamma_{SS}$:
\begin{equation}
h(b)=\sum_{n\geq 1}\bigg(T_fN_f\frac{\alpha_s}{3\pi}\bigg)^{n-2}\bigg[h_n
+O\bigg(\frac{1}{N_f}\bigg)\bigg]
\end{equation}
and $g_n$ are the numbers that are generated through the  
$\epsilon$-expansion of the following identity
\begin{equation}
\sum_{n}g_n\epsilon^{n-1}=
\bigg[4-\sum_{n>1}\bigg(\frac{3}{2^n}+\frac{n}{2}\bigg)\epsilon^{n-2}\bigg]
~{\rm exp}\bigg(\sum_{l>2}
\frac{2^l-3-(-1)^l}{l}\zeta(l)\epsilon^l\bigg)~~~~~. 
\end{equation}

\section{Unravelling  the $\delta=1$ IR renormalon knot}

The contributions of both UV and IR renormalons to $\Pi_{S}$ 
is given by the corresponding expression for     
$G(\delta)$ \cite{Broadhurst:2001yc}:
\begin{equation}
G_D(\delta)=\frac{2}{1-\delta}-\frac{1}{2-\delta}+\frac{8(1-\delta)}{3}
\sum_{k=2}^{\infty}\frac{(-1)^k k}{(k^2-(1-\delta)^2)^2}
\end{equation}
It can be separated into two pieces as $G_D(\delta)=G_{-}(\delta)+G_{+}(\delta)$. The Borel resummable UV renormalons are contained in the first term
\begin{equation}
G_{-}(\delta)=-\frac{2}{3}\sum_{k>0}\frac{(-1)^k}{(k+\delta)}~~~,
\end{equation}
with singularities located at $\delta<0$. 
At the  large-$N_f$, limit it generates the 
sign-constant asymptotic series. After performing a 
naive nonabelianization, namely after the replacement $b\rightarrow -\beta_0\alpha_s(\mu^2)/\pi$, one obtains 
sign-alternating asymptotic perturbative QCD series 
for  $\Pi_S$.

The IR renormalons are defined by $G_{+}(\delta)$, which has the 
following expression
\begin{equation}
G_{+}(\delta)=\frac{2}{1-\delta}-\frac{1}{2-\delta}+\frac{2}{3}\sum_{k>2}
\frac{(-1)^k}{(k-\delta)^2}~~~~~~.
\label{IRR}
\end{equation}
Contary to the vector case (see e.g. Ref.~\cite{Broadhurst:1993si})
one can observe the appearance in Eq.~(\ref{IRR}) 
of the first IR renormalon at $\delta=1$. 
In accordance with the general rules of renormalon calculus,
this IR renormalon  generates the   
$\Lambda^2/Q^2$ contribution to the   $\tilde{D}_S(Q^2)$-function of Eq.~(6).
It should be stressed
that this term cannot appear within the  operator-product expansion machinery. 
Indeed,  the first higher-twist operator, contributing 
to the $\tilde{D}_S(Q^2)$-function in the massless scheme comes from the 
gluon condensate $\langle (\alpha_s/\pi) G^2\rangle_0$ \cite{Shifman:1979bx}, 
related to the IR renormalon pole at $\delta=2$ (note that the perturbative 
QCD correction to the gluon condensate  was calculated in 
Ref.~\cite{Surguladze:1990sp}). 

The reason for the appearance of the $\delta=1$ IR renormalon contribution to 
the $\tilde{D}_S$-function is related to the fact 
that the dispersion relation 
of Eq.~(6) is ill-defined. Indeed, contrary to the vector case, where 
the Ward identity allows us to fix the expression for $\Pi_V(Q^2)$ at 
zero transferred momentum, 
the  expression for $\Pi_S(0)$ is infinite. Therefore, the 
dispersion relation for the first derivative of $\Pi_S(Q^2)/Q^2$ of Eq.~(6) 
contains an ambiguity of order $\Lambda^2/Q^2$. For large  
momentum transfer this problem can be 
neglected. However, one should avoid the 
application of Eq.~(6) in the process of long-distance studies, 
e.g.  the determination  of light-quark masses values. 
In this case it is necessary to 
use the dispersion relation for the second derivative of $\Pi_S$
(see Eq.~(7)), which is free from unphysical $\Lambda^2/Q^2$ term.
This renormalon rediscovery of the advantages of the twice-differentiated 
Euclidean construct of Eq.~(7), which was originally used in Ref.
\cite{Becchi:1981vz} and later on in Refs.~
\cite{Chetyrkin:1995qu}--\cite{Colangelo:1997uy}, 
is setting its application for the extraction 
of light-quark masses on a more solid background. It should  also be 
mentioned, that the spectral density $R_S(s)$ of both Eq.~(6) and Eq.~(7) is 
also free from contributions of $\delta=1$ IR renormalon. Moreover, since 
the factor $\pi\delta/\rm sin (\pi\delta)$ of analytic continuation of the 
Euclidean  constructs  to the Minkowskian region is removing all 
single poles in Eq.~(15) and transforming double poles in $\delta$ into 
single poles, the Borel image of  $R_S(s)$ contains  $\delta>2$
IR renormalons only, which are associated with the contributions to $R_S(s)$ 
of the operators $O_k$ with dimension $d_k\geq 6$.

\section{Correlator of scalar quark currents and \\ large-$\beta_0$ 
expansion} 

The large-$\beta_0$ expansion in the $\overline{\rm MS}$ scheme  is 
widely used to generate perturbative series from the direct calculations 
of the renormalon-chain diagrams, which contain the information on 
both UV and IR renormalons structure of the related Borel transforms. 
Let us summarize the results 
of  
its application for modelling the behaviour of the 
perturbative expansions 
for basic quantities, related to the correlators of the scalar and pseudoscalar
quark currents (see  Ref.~ \cite{Broadhurst:2001yc}).
The quantities defined above  in Eqs.~(6),(7) have the following expansions:
\begin{eqnarray}
\tilde{D}_S(Q^2)&=& 3[m(Q^2)]^2\left(1+\sum_{n>0}d_n a^n\right) \\
\overline{D}_S(Q^2)&=&3[m(Q^2)]^2\left(1+\sum_{n>0}\overline{d}_n a^n\right) 
\\ 
R_S(s)&=&3[m(s)]^2\left(1+\sum_{n>0}s_n a^n\right)~~.
\end{eqnarray} 
In leading order of the large-$\beta_0$ expansion, the coefficients 
of these series, obtained using the naive nonabelianization procedure, 
can be presented as
\begin{eqnarray}
d_n^{NNA}&=&\beta_0^{n-1}\tilde{\Delta}_{n+1} \\    
s_n^{NNA}&=&\beta_0^{n-1}\Delta_{n+1} \\
\overline{d}_{n}^{NNA}&=&\beta_0^{n-1}\overline{\Delta}_{n+1}
\end{eqnarray}
where the coefficients $\tilde{\Delta}_{n+1}$, $\Delta_{n+1}$ 
and $\overline{\Delta}_{n+1}$ were calculated in Ref.~\cite{Broadhurst:2001yc}
in the $\overline{\rm MS}$ scheme. The concrete results of calculations 
are presented in Table 1. 
%\newpage
\begin{center}{\bf Table~1:} The values of coefficients in Eqs.~(19),(20),(21).
\end{center}$$\begin{array}{|l|rrr|}\hline n&
\widetilde{\Delta}_n\quad{}&\Delta_n\quad{}&\overline{\Delta}_n\quad{}\\\hline
2 &  7.33333 &  7.33333 &  5.33333\\
3 &  10.4696 &  7.17968 &  3.13622\\
4 &   32.6145 &  8.48885 &  11.6754\\
5 &    97.9534 &  4.36402 &  0.10978\\
6 &  503.887 &  2.96849 &  112.074\\
7 &  2194.28 & -54.2101 & -325.157\\
8 &   16465.8 &  123.639 &  3300.11\\
\hline\end{array}$$
One can see that the coefficients $\tilde{\Delta}_{n+1}$ are increasing 
very fast. This feature is related  
to the manifestation of $\delta=1$ IR renormalon 
contribution to Eq.~(13). The suppression of this singularity in 
the expressions 
for $R_S(s)$ and $\overline{D}_S(Q^2)$ leads to a rather well-behaved 
perturbative 
series up to $n=6$-loop order  for $R_S(s)$ and $n=5$-loop order
in the latter case, where the corresponding 
coefficient $\Delta_5$ is remarkably small.  The factorial growth of the 
related series starts to manifest itself at the 
level of  higher-order corrections. 

Consider now the uncertainties of the large-$\beta_0$
series, comparing the numerical values  for the coefficients 
of Eqs.~(19),(20),(21) with the explicit numbers, obtained 
as the result of analytical calculations, performed in 
Refs.~\cite{Gorishnii:1984cu}--
\cite{Chetyrkin:1997sr} (see Table ~2).
%\newpage
\begin{center}{\bf Table~2:}
Ratios of NNA estimates to exact results.
\end{center}
$$\begin{array}{|l|r|r|r|}
 \hline
 Coeff.    &                   N_f=3 &    N_f=4 &    N_f=5\\
 \hline
 d^{\rm NNA}_2/d_2                       & 0.514 & 0.496 & 0.477\\
 s^{\rm NNA}_2/s_2                       & 0.507 & 0.490 & 0.472\\
 \overline{d}^{\rm NNA}_2/\overline{d}_2 & 0.498 & 0.484 & 0.469\\
 \hline
 d^{\rm NNA}_3/d_3                       & 0.354 & 0.346 & 0.339\\
 s^{\rm NNA}_3/s_3                       & 0.482 & 0.565 & 0.747\\
 \overline{d}^{\rm NNA}_3/\overline{d}_3 & 0.764 & 0.871 & 1.081\\
 \hline
\end{array}$$
One can see that  contrary to the applications 
of the scheme-invariant approach for estimates of higher-order 
perturbative QCD corrections in the scalar channel \cite{Chetyrkin:1997wm},
the large-$\beta_0$ approximation gives the  correct signs  
for the coefficients of all considered quantities. The reason for the 
failure of the scheme-invariant estimates at the 4-loop level 
will be clarified below. It should also be  stressed that 
although the large-$\beta_0$  predictions are of the same 
order of magnitude as the exactly calculable coefficients, the 
application of the naive nonabelianization procedure in the scalar channel 
has the tendency to underestimate the concrete perturbative terms by a 
factor of order 0.5. Keeping this in mind, and taking into account 
the results from Table~1, we arrive at the conclusion that the 
large-$\beta_0$ approximation indicates, that 
the  5-loop corrections to  the perturbative series for the 
$\overline{D}_S$ function and $R_S(s)$, which enter in the determination 
of the strange-quark mass  within QCD sum rules 
method (see Refs.~\cite{Becchi:1981vz},\cite{Chetyrkin:1995qu}-
\cite{Colangelo:1997uy}), are extremely small. Indeed, 
fixing $N_f=3$ and $\alpha_s\approx 0.3$ we obtain 
\begin{eqnarray}
\overline{D}_S&=&1+3.67\left(\frac{\alpha_s}{\pi}\right)+
14.17\left(\frac{\alpha_s}{\pi}\right)^2+77.36\left(\frac{\alpha_s}{\pi}\right)^3+2\times1.26\left(\frac{\alpha_s}{\pi}\right)^4 \\ \nonumber
&=&1+0.350 + 0.129 + 0.067 + 0.0002~~~
\end{eqnarray}
and 
\begin{eqnarray}
R_S(s)&=&1+5.667\left(\frac{\alpha_s}{\pi}\right)
+31.864\left(\frac{\alpha_s}{\pi}\right)^2
+89.156\left(\frac{\alpha_s}{\pi}\right)^3
+98\left(\frac{\alpha_s}{\pi}\right)^4 \\ \nonumber 
&=&1 + 0.541 + 0.291 + 0.078 + 0.008~~~. 
\end{eqnarray}
Therefore,  large-$\beta_0$ estimates indicate  that the doubts
about a  serious underestimate 
of the errors of the QCD strange-quark-mass  
extraction from the pseudoscalar 
correlator at the $(\alpha_s^2)$--order \cite{Chetyrkin:1995qu}, 
formulated  in Ref.~\cite{Yndurain:1998gk} 
using the  positivity of a  spectral function for a pseudoscalar correlator,   cannot be related to the non-applicability of the perturbative QCD expansions
at rather small energy scale. 
Moreover, the  $O(\alpha_s^3)$ result of Ref.~\cite{Chetyrkin:1997xa}
$\overline{m}_{s}(1~\rm GeV)=205 \pm 19 MeV$ is closer 
to the upper bound from Ref.~\cite{Yndurain:1998gk}, 
namely  $140 \leq \overline{m}_{s}(1~\rm GeV)\leq 254$ MeV.
At the energy scale of  2~GeV the results of Ref.~\cite{Chetyrkin:1997xa}
are equivalent to the value $\overline{m}_{s}(2~\rm GeV)\sim$ 148 MeV, 
which are higher than the current lattice results (for a review, 
see Ref.~\cite{Gupta:2001cu}). Note, however, that the current determination 
of light-quark masses from QCD sum rules for the scalar and pseudoscalar 
two-point functions does  not touch the problem of the 
possible manifestation    
of the  instanton contributions  in  the theoretical part of these sum rules 
(see Ref.~\cite{Gabrielli:1993wv}). This is why  the determination of 
the {\it s}-quark mass in Cabibbo-suppressed $\tau$-lepton 
decay characteristics, which seem to be free from instanton  
contributions,  attracts more and more attention (see e.g. Ref.~\cite{Chetyrkin:1998ej} and in particular Ref.~\cite{Korner:2000wd}). 
Note that the central value of the 
{\it s}-quark mass  
result of  Ref.~\cite{Korner:2000wd}, namely  
$\overline{m}_{s}(1~\rm GeV)=176 \pm 37_{exp}\pm 13_{th}$ MeV, 
which is equivalent to  
$\overline{m}_{s}(2~\rm GeV)\sim$ 128 MeV,
is lower  than of  
the one of Ref.~\cite{Chetyrkin:1997xa} and is closer 
to the lattice results, summarized in the review of Ref.~\cite{Gupta:2001cu}. 
One of the reasons can be  related  
to 
the approach  applied in  Ref.~\cite{Korner:2000wd},  of 
resummation of the infinite subset of  $\pi^2$-dependent effects 
that arize from 
analytical continuation of the mass-dependent quantities  
from  the Euclidean to the Minkowskian region. 
Within the large-$\beta_0$ approximation 
the special features of this procedure  
were  studied in Ref.~\cite{Broadhurst:2001yc}.

\section{Pole vs  running quark masses and \\the large-$\beta_0$ 
approximation} 

Another example where the naive nonabelianization 
procedure \cite{Broadhurst:1995se} and the large-$\beta_0$ 
estimates are  working reasonably well, is the 
relation between pole and running quark masses.
Indeed, this relation was recently evaluated analytically 
at the $\alpha_s^3$-order \cite{Melnikov:2000qh}. The results 
of these calculations, which confirm those obtained in 
Ref.~\cite{Chetyrkin:1999ys} in a  semi-analytical way, have the 
following numerical form \cite{Melnikov:2000qh}:
\begin{eqnarray}
M&=&\overline{m}(\overline{m})\bigg[1+\frac{4}{3}
\left(\frac{\overline{\alpha}_s}{\pi}\right)
+\left(\frac{\overline{\alpha}_s}{\pi}\right)^2 (-1.0414 N_f+13.4434) \\ \nonumber 
&+& \left(\frac{\overline{\alpha}_s}{\pi}\right)^3(0.6527 N_f^2-26.655 N_f + 190.595)\bigg]
\end{eqnarray}
with  $\overline{\alpha}_s=\overline{\alpha}_s(\overline{m})$. 
This series can be rewritten in 
terms of the scheme-independent coefficients of the QCD 
$\beta$-function as \cite{Melnikov:2000qh}:
\begin{eqnarray}
M&=&\overline{m}(\overline{m})\bigg[1+\frac{4}{3}
\left(\frac{\overline{\alpha}_s}{\pi}\right)
+\left(\frac{\overline{\alpha}_s}{\pi}\right)^2 (6.248 \beta_0 - 3.739) 
\label{betp}
\\ \nonumber 
&+& \left(\frac{\overline{\alpha}_s}{\pi}\right)^3(23.497 \beta_0^2+6.248\beta_1+1.019 \beta_0- 29.94)\bigg]~~~,
\end{eqnarray}
where $\beta_0$ is as  defined above and $\beta_1=[102-(38/3)N_f]/16$.
It is possible now to  deduce that the large-$\beta_0$ approximation 
works reasonably well in this case as well, thanks to  the 
cancellations between the term  proportional to the  $\beta_1$ coefficient 
and the conformal $\beta$-independent one,  
in the order  correction of order $\alpha_s^3$ to  Eq.~(\ref{betp})
the large--$\beta_0$ approximation is working reasonably  well     
\footnote{I wish to thank E.~Gardi for  discussions on  questions  
related to this topic.}. This feature is also illustrated  
in Table 3, taken from  Ref.~\cite{Chetyrkin:1999ys}, 
where the comparison of the
results of semi-analytic calculations of the $O(\alpha_s^3)$ correction 
to Eq.~(24) \cite{Chetyrkin:1999ys}, presented in the second column,
are compared with the estimates of these terms, made in 
Ref.~\cite{Chetyrkin:1997wm}  
using  the scheme-invariant methods (namely  the effective-charges 
approach (ECH) of Ref.~\cite{Grunberg:1984fw} and  the PMS approach of 
Ref.~\cite{Stevenson:1981vj}) and  the large-$\beta_0$ estimates 
of Ref.~\cite{Beneke:1995qe}.  
\begin{center}{\bf Table~3:} Comparison
        of the results of Ref.~\cite{Chetyrkin:1999ys}
        paper with estimates based on  ECH, PMS and the large-$\beta_0$
        approximation for $M/\overline{m}(\overline{m})$.
\end{center}      
\begin{center}
\begin{tabular}{|l|r|r|r|r|}
\hline        
 $N_f$ 
        & \cite{Chetyrkin:1999ys}
        & \cite{Chetyrkin:1997wm} (ECH)
        & \cite{Chetyrkin:1997wm} (PMS)
        & \cite{Beneke:1995qe} (large-$\beta_0$)
        \\
        \hline
        $2$ &
        $    143(3)$
        & $152.71$ & $153.76$ & $137.23$\\
        $3$ &
        $    119(3)$
        & $124.10$ & $124.89$ & $118.95$\\
        $4$ &
        $     96(2)$
        & $97.729$ & $98.259$ & $101.98$\\
        $5$ &
        $     75(2)$
        & $73.616$ & $73.903$ & $86.318$\\
\hline
\end{tabular}
\end{center}
One can see that  the scheme-invariant methods, 
developed in other cases in Ref.~\cite{Kataev:1995vh}, also give 
reasonable estimates. However, there are  cases, where  
the comparison of the results of estimates, produced by 
scheme-invariant methods or the large $\beta_0$-approach, with the explicitly 
calculated corrections of the perturbative series under consideration, 
demonstrate  larger ambiguities.  

\section{Limitations  of  estimating procedures: when and why? }  

It should be mentioned that unlike the case of  the qualitative success 
of the large-$\beta_0$ estimates in Ref.~\cite{Broadhurst:2001yc}
of the coefficients of the perturbative series of   
quantities related to the Green 
function of the scalar and pseudoscalar quark currents, 
the application of the scheme-invariant estimates  in this case 
result in the appearance of large and negative- 
order $\alpha_s^4$ corrections in the expression 
for $R_S(s)$ \cite{Chetyrkin:1997wm}.  
It is thus  worth-while trying   
to understand the reason  for  the appearance  of these 
suspicious numbers and, therefore, the limitations of the applications
of the scheme-invariant 
procedure of Ref.~\cite{Kataev:1995vh}, used in Ref.~\cite{Chetyrkin:1997wm}.
These studies were done in Ref.~\cite{Broadhurst:2001yc} and their main 
conclusions will be summarized here.

It is worth-while recalling that the approach of Ref.~\cite{Kataev:1995vh} 
is constructed for the 
estimates of higher-order perturbative corrections to renormalization-
group-invariant quantities of Eq.~(2), which obey the renormalization-group 
equation without anomalous dimension term, with the $\beta$-function
defined as $\beta(a)=-\sum_{i\geq 0} \beta_{i}a^{i+2}=-\beta_0a^2 
(1+\sum_{i\geq 1}c_i a^{i})$, where $c_i=\beta_i/\beta_0$.
It is known that the method  of the effective charges of 
Ref.~\cite{Grunberg:1984fw} prescribes to define the scheme where 
all higher-order corrections to $D(a)$ will be nullified and the 
corresponding expression will have the following form: $D(\tilde{a})=d_0
\tilde{a}$. The new expansion parameter  obeys the renormalization-group 
equation with  the effective-charges $\beta$-function, defined as 
\begin{equation}
\tilde{\beta}(\tilde{a})=
-\sum_{n\geq 0} \tilde{\beta}_{n}\tilde{a}^{i+2}~~,
\end{equation}
where the coefficients $\tilde{\beta}_n$ are scheme-invariant. 
The basic formula for generating the estimates from the scheme-invariant 
procedure of Ref.~\cite{Kataev:1995vh} is:
\begin{equation}
\frac{\tilde{\beta}_n-\beta_n}{(n-1)}=\beta_0(d_n-\Omega_n)~~,
\label{om}
\end{equation}
where $\Omega_n$ are defined as 
\begin{eqnarray}
\Omega_2&=&d_1(c_1+d_1) \\ \nonumber 
\Omega_3&=&d_1\bigg(c_2-\frac{1}{2}c_1d_1-2d_1^2+3d_2\bigg)\\ \nonumber
\Omega_4&=&d_1\bigg(c_3-\frac{4}{3}c_2d_1+\frac{2}{3}c_1d_2+
\frac{14}{3}d_1^3-\frac{28}{3}d_1d_2+4d_3\bigg) \nonumber\\
&&{}
+\frac{1}{3}\bigg(c_2d_2-c_1d_3+5d_2^2\bigg)~~~.
\label{om4}
\end{eqnarray}        
at 5 loops. The idea that lies  beyond the procedure of 
Ref.~\cite{Kataev:1995vh} is 
that reasonable estimates of the 
higher-order coefficients $d_n$ of the perturbative series for  physical 
quantities can be obtained   
if the coefficients $\tilde{\beta}_n$ of the process- 
dependent, effective-charges $\beta$-functions are of the same sign and 
magnitude as  the coefficients $\beta_n$ of the $\beta$-finction in the 
$\overline{\rm MS}$ scheme. 
This assumption turns out to be true  at the 4-loop level 
in QCD for the perturbative series of  
deep-inelastic scattering sum rules and the $e^+e^-$-annihilation 
$D$-function, 
in the cases $N_f=4$ and $N_f=3$ in particular. 
Indeed   the numerical analogues of the analytical 
expressions  for the 4-loop  
QCD $\beta$-function \cite{vanRitbergen:1997va} have the following form 
  \begin{eqnarray}
\beta(N_f=3)&=&-2.250a^2-4.000a^3          -10.060a^4-47.228a^5
\label{bet3}\\
\beta(N_f=4)&=&-2.083a^2-3.208a_s^3-\phantom{1}6.349a^4-31.387a^5
\label{bet4}\\
\beta(N_f=5)&=&-1.917a^2-2.417a^3-\phantom{1}2.827a^4-18.852a^5
\label{bet5}
\end{eqnarray}
with $a\equiv\alpha_s/\pi$. It should be stressed that at $N_f=3$ 
and $N_f=4$ the effective-charges $\beta$-functions of the 
polarized and unpolarized Bjorken sum rules and of the 
$e^+e^-$-annihilation $D$-function have a similar 4-loop  behaviour, 
provided  the explicitly unknown coefficients $d_3$ in 
$\Omega_4$ are modelled by the numerical expressions for $\Omega_3$.
However, this is not true in the case  of renormalization-group-invariant 
quantities, related to the correlators of scalar and pseudoscalar 
currents. Indeed, for    
the renormalization-group-invariant quantity related to the 
$\overline{D}_s$-function of Eq.~(7) as 
\begin{equation}
\overline{R}_D(Q^2)=-\frac{1}{2}\frac{d\rm log \overline{D}_S(Q^2)}
{d\rm log Q^2}
\end{equation}    
the behaviour of the   effective-charges $\beta$-function  
in the  $O(\tilde{a}^3)$ and $O(\tilde{a}^3)$-approximations 
differs significantly from the behaviour of Eqs.~(30),(31),(32) 
\cite{Broadhurst:2001yc}.  The coresponding  expressions  
\begin{eqnarray}
\widetilde{\beta}(N_f=3)&=&-2.250\tilde{a}^2
-4.000\tilde{a}^3+58.920\tilde{a}^4-2148.503\tilde{a}^5
\label{beff3}\\
\widetilde{\beta}(N_f=4)&=&-2.083\tilde{a}^2
-3.208\tilde{a}^3+53.852\tilde{a}^4-1687.191\tilde{a}^5
\label{beff4}\\
\widetilde{\beta}(N_f=5)&=&-1.917\tilde{a}^2
-2.417\tilde{a}^3+49.356\tilde{a}^4-1303.490\tilde{a}^5
\label{beff5}
\end{eqnarray}
indicate the appearance of a spurious perturbative infrared 
fixed point. This spurious zero is compensated by the appearance 
of large and negative $\tilde{a}^5$ contributions.
In the case of the $R_D$-function, similar features were 
already observed at the 3-loop 
level in Ref.~\cite{Gorishnii:1991zr} and at the 4-loop 
in Ref.~\cite{Vermaseren:1997fq} both in the Minkowskian region; 
they were  studied in detail in the Euclidean region 
in Ref.~\cite{Gardi:1998rf}.
Here we conclude that these features demonstrate the limitations 
of the application of the scheme-invariant methods to the  
estimates  
 of the O($\alpha_s^4$)-corrections to the $\tilde{D}_S$ function
of Eq.~(6), the  $\overline{D}_S$ function of Eq.~(7), and their spectral 
density $R_S(s)$. 

However, the approximation  of the  
large $\beta_0$-expansion, which we think  
is working reasonably well in the case of the  correlator of the scalar and 
pseudoscalar quark currents \cite{Broadhurst:2001yc},  also has definite 
limitations. The inapplicability of this method for the estimates 
of higher-order corrections to the DGLAP non-singlet kernel, 
studied in detail in 
Ref.~\cite{Mikhailov:1998xi}, can be explained by the absence of  
renormalon-type contributions to the anomalous dimensions of 
the non-singlet operators. A more interesting fact is related to 
the contradiction between the  large-$\beta_0$ 
estimates  of the high-order  
perturbative corrections to the coefficient functions 
of $N=2$ and $N=4$ non-singlet Mellin moments, for the  $F_2$ structure 
function \cite{Mankiewicz:1997gz}, and the result of explicit 
calculations performed in Ref.~\cite{Larin:1994vu}. The limitations 
of the predictive possibilities of the large-$\beta_0$ method 
in this case are still unclear. 
  
{\bf Acknowledgements}

I am grateful to  D.J. Broadhurst and C.J. Maxwell for the 
productive and pleasant  
collaboration, which resulted in the publication of 
Ref.~\cite{Broadhurst:2001yc}. Different aspects 
of this work were discussed  at the Quarks-2000 International 
Seminar in the talks of D.J.B. and A.L.K., and are  described in part above. 
It is a pleasure to thank G. Altarelli, 
V.M. Braun, J. Ellis, J. Fischer, N.V. Krasnikov, S.A. Larin, 
N. Paver, A.A. Pivovarov, 
A. Schafer  and F.J. Yndurain for 
discussions and constructive questions and  comments.   

Part of the work on the problems approached in this paper 
was done in Russia within 
the framework of scientific projects 
of RFBR Grants N 99-01-00091, N 00-02-1742.


\begin{thebibliography}{1}
\bibitem{Vainshtein:1978nn}
A.~I.~Vainshtein, M.~B.~Voloshin, V.~I.~Zakharov, V.~A.~Novikov, L.~B.~Okun and M.~A.~Shifman,
%``Sum Rules For Light Quarks In Quantum Chromodynamics. (In Russian),''
Sov.\ J.\ Nucl.\ Phys.\ {\bf 27} (1978) 274.
%%CITATION = SJNCA,27,274;%%
\bibitem{Becchi:1981vz}
C.~Becchi, S.~Narison, E.~de Rafael and F.~J.~Yndurain,
%``Light Quark Masses In Quantum Chromodynamics And Chiral Symmetry Breaking,''
Z.\ Phys.\ {\bf C8}, 335 (1981).
%%CITATION = ZEPYA,C8,335;%%
\bibitem{Kataev:1983xu}
A.~L.~Kataev, N.~V.~Krasnikov and A.~A.~Pivovarov,
%``The Use Of The Finite Energetic Sum Rules For The Calculation Of The Light Quark Masses,''
Phys.\ Lett.\ {\bf B123} (1983) 93.
%%CITATION = PHLTA,B123,93;%%
%\cite{Narison:1983af}
\bibitem{Narison:1983af}
S.~Narison, N.~Paver, E.~de Rafael and D.~Treleani,
%``Light Quark Mass Differences In Quantum Chromodynamics,''
Nucl.\ Phys.\ B {\bf 212} (1983) 365.
%%CITATION = NUPHA,B212,365;%%
\bibitem{Gorishnii:1984zi}
S.~G.~Gorishny, A.~L.~Kataev and S.~A.~Larin,
%``Next Next-To-Leading Perturbative QCD Corrections And Light Quark Masses,''
Phys.\ Lett.\ {\bf B135} (1984) 457.
%%CITATION = PHLTA,B135,457;%%
%\cite{Dominguez:1987aa}
\bibitem{Dominguez:1987aa}
C.~A.~Dominguez and E.~de Rafael,
%``Light Quark Masses In QCD From Local Duality,''
Ann. \ Phys.\ {\bf 174} (1987) 372.
%%CITATION = APNYA,174,372;%%
\bibitem{Gasser:1982ap}
J.~Gasser and H.~Leutwyler,
%``Quark Masses,''
Phys.\ Rep.\ {\bf 87} (1982) 77.
%%CITATION = PRPLC,87,77;%%
\bibitem{Kataev:1985vh}
A.~L.~Kataev, N.~V.~Krasnikov and A.~A.~Pivovarov,
%``Finite Energy Sum Rules And Dynamical Properties Of Hadrons In QCD,''
Preprint INR-P-0389 (1984);\\ 
in Proc. Quarks-84 Int. Seminar, Tbilisi, 1984.
\bibitem{Narison:1987da}
S.~Narison,
%``Chiral Symmetry Breaking And The Light Meson Systems,''
Riv.\ Nuovo Cim.\ {\bf 10} (1987) 1.
%%CITATION = RNCIB,10,1;%%
\bibitem{Narison:2000uj}
S.~Narison,
%``Light quark masses 99,''
Nucl.\ Phys.\ Proc.\ Suppl.\ {\bf 86} (2000) 242
[hep-ph/9911454].
%%CITATION = HEP-PH 9911454;%%
\bibitem{Gupta:2001cu}
R.~Gupta and K.~Maltman,
%``Light quark masses: a status report at DPF 2000,''
hep-ph/0101132.
%%CITATION = HEP-PH 0101132;%%
%\cite{vanRitbergen:1997va}
\bibitem{vanRitbergen:1997va}
T.~van Ritbergen, J.~A.~Vermaseren and S.~A.~Larin,
%``The four-loop beta function in quantum chromodynamics,''
Phys.\ Lett.\ {\bf B400} (1997) 379
[hep-ph/9701390].
%%CITATION = HEP-PH 9701390;%%
%\cite{Chetyrkin:1997dh}
\bibitem{Chetyrkin:1997dh}
K.~G.~Chetyrkin,
%``Quark mass anomalous dimension to O(alpha(s)**4),''
Phys.\ Lett.\ {\bf B404} (1997) 161
[hep-ph/9703278].
%%CITATION = HEP-PH 9703278;%%
%\cite{Vermaseren:1997fq}
\bibitem{Vermaseren:1997fq}
J.~A.~Vermaseren, S.~A.~Larin and T.~van Ritbergen,
%``The 4-loop quark mass anomalous dimension and the invariant quark  mass,''
Phys.\ Lett.\ {\bf B405} (1997) 327
[hep-ph/9703284].
%%CITATION = HEP-PH 9703284;%%
%\cite{Gray:1990yh}
\bibitem{Gray:1990yh}
N.~Gray, D.~J.~Broadhurst, W.~Grafe and K.~Schilcher,
%``Three Loop Relation Of Quark (Modified) Ms And Pole Masses,''
Z.\ Phys.\ {\bf C48} (1990) 673.
%%CITATION = ZEPYA,C48,673;%%
%\cite{Fleischer:1999dw}
\bibitem{Fleischer:1999dw}
J.~Fleischer, F.~Jegerlehner, O.~V.~Tarasov and O.~L.~Veretin,
%``Two-loop {QCD} corrections of the massive fermion propagator,''
Nucl.\ Phys.\ {\bf B539} (1999) 671
[hep-ph/9803493].
%%CITATION = HEP-PH 9803493;%%
%\cite{Beneke:1995qe}
\bibitem{Beneke:1995qe}
M.~Beneke and V.~M.~Braun,
%``Naive nonAbelianization and resummation of fermion bubble chains,''
Phys.\ Lett.\ {\bf B348} (1995) 513
[hep-ph/9411229]; \\
%%CITATION = HEP-PH 9411229;%%
%\cite{Ball:1995ni}
%\bibitem{Ball:1995ni}
P.~Ball, M.~Beneke and V.~M.~Braun,
%``Resummation of (beta0 alpha-s)**n corrections in QCD: Techniques and applications to the tau hadronic width and the heavy quark pole mass,''
Nucl.\ Phys.\ {\bf B452} (1995) 563
[hep-ph/9502300].
%%CITATION = HEP-PH 9502300;%%
%\cite{Philippides:1995jw}
\bibitem{Philippides:1995jw}
K.~Philippides and A.~Sirlin,
%``Leading vacuum polarization contributions to the relation between pole and running masses,''
Nucl.\ Phys.\ {\bf B450} (1995) 3
[hep-ph/9503434].
%%CITATION = HEP-PH 9503434;%%
%\cite{Chetyrkin:1997wm}
\bibitem{Chetyrkin:1997wm}
K.~G.~Chetyrkin, B.~A.~Kniehl and A.~Sirlin,
%``Estimations of order alpha(s)**3 and alpha(s)**4 corrections to  mass-dependent observables,''
Phys.\ Lett.\ {\bf B402} (1997) 359
[hep-ph/9703226].
%%CITATION = HEP-PH 9703226;%%
%\cite{Kataev:1995vh}
\bibitem{Kataev:1995vh}
A.~L.~Kataev and V.~V.~Starshenko,
%``Estimates of the higher order QCD corrections to R(s), R(tau) and deep inelastic scattering sum rules,''
Mod.\ Phys.\ Lett.\ {\bf A10} (1995) 235
[hep-ph/9502348].\\
%%CITATION = HEP-PH 9502348;%%
A.~L.~Kataev and V.~V.~Starshenko,
%``The renormalization group inspired approaches and estimates of the tenth order corrections to the muon anomaly in QED,''
Phys.\ Rev.\ D {\bf 52} (1995) 402
[hep-ph/9412305].
%%CITATION = HEP-PH 9412305;%%
%\cite{Chetyrkin:1999ys}
\bibitem{Chetyrkin:1999ys}
K.~G.~Chetyrkin and M.~Steinhauser,
%``Short distance mass of a heavy quark at order alpha(s)**3,''
Phys.\ Rev.\ Lett.\ {\bf 83} (1999) 4001
[hep-ph/9907509].
%%CITATION = HEP-PH 9907509;%%
%\cite{Melnikov:2000qh}
\bibitem{Melnikov:2000qh}
K.~Melnikov and T.~v.~Ritbergen,
%``The three-loop relation between the MS-bar and the pole quark masses,''
Phys.\ Lett.\ {\bf B482} (2000) 99
[hep-ph/9912391].
%%CITATION = HEP-PH 9912391;%%
%\cite{Bijnens:1995ci}
\bibitem{Bijnens:1995ci}
J.~Bijnens, J.~Prades and E.~de Rafael,
%``Light quark masses in QCD,''
Phys.\ Lett.\ {\bf B348} (1995) 226
[hep-ph/9411285].
%%CITATION = HEP-PH 9411285;%%
%\cite{Chetyrkin:1995qu}
\bibitem{Chetyrkin:1995qu}
K.~G.~Chetyrkin, C.~A.~Dominguez, D.~Pirjol and K.~Schilcher,
%``Mass singularities in light quark correlators: The Strange quark case,''
Phys.\ Rev.\ {\bf D 51} (1995) 5090
[hep-ph/9409371].
%%CITATION = HEP-PH 9409371;%%
%\cite{Jamin:1995vr}
\bibitem{Jamin:1995vr}
M.~Jamin and M.~Munz,
%``The Strange quark mass from QCD sum rules,''
Z.\ Phys.\ {\bf C66} (1995) 633
[hep-ph/9409335].
%%CITATION = HEP-PH 9409335;%%
%\cite{Chetyrkin:1997xa}
\bibitem{Chetyrkin:1997xa}
K.~G.~Chetyrkin, D.~Pirjol and K.~Schilcher,
%``Order-alpha(s)**3 determination of the strange quark mass,''
Phys.\ Lett.\ {\bf B404} (1997) 337
[hep-ph/9612394].
%%CITATION = HEP-PH 9612394;%%
%\cite{Colangelo:1997uy}
\bibitem{Colangelo:1997uy}
P.~Colangelo, F.~De Fazio, G.~Nardulli and N.~Paver,
%``On the QCD sum rule determination of the strange quark mass,''
Phys.\ Lett.\ B {\bf 408} (1997) 340
[hep-ph/9704249].
%%CITATION = HEP-PH 9704249;%%
%\cite{Broadhurst:2001yc}
\bibitem{Broadhurst:2001yc}
D.~J.~Broadhurst, A.~L.~Kataev and C.~J.~Maxwell,
%``Renormalons and multiloop estimates in scalar correlators, Higgs decay  and quark-mass sum rule,''
Nucl.\ Phys.\ {\bf B592} (2001) 247
[hep-ph/0007152].
%%CITATION = HEP-PH 0007152;%%
%\cite{Bigi:1994em}
\bibitem{Bigi:1994em}
I.~I.~Bigi, M.~A.~Shifman, N.~G.~Uraltsev and A.~I.~Vainshtein,
%``The Pole mass of the heavy quark. Perturbation theory and beyond,''
Phys.\ Rev.\ D {\bf 50} (1994) 2234
[hep-ph/9402360].
%%CITATION = HEP-PH 9402360;%%
%\cite{Beneke:1994sw}
\bibitem{Beneke:1994sw}
M.~Beneke and V.~M.~Braun,
%``Heavy quark effective theory beyond perturbation theory: Renormalons, the pole mass and the residual mass term,''
Nucl.\ Phys.\ {\bf B426} (1994) 301
[hep-ph/9402364].
%%CITATION = HEP-PH 9402364;%%
\bibitem{Altarelli}
G. Altarelli, CERN-TH/95-309; in  Proceedings of Int. School 
of Subnuclear Physics, July 1995, Erice, Italy.
\bibitem{Beneke:2000kc}
M.~Beneke,
%``Renormalons,''
Phys.\ Rep.\ {\bf 317} (1999) 1
[hep-ph/9807443];\\
%%CITATION = HEP-PH 9807443;%%
M.~Beneke and V.~M.~Braun,
%``Renormalons and power corrections,''
hep-ph/0010208.
%%CITATION = HEP-PH 0010208;%%
%\cite{Lautrup:1977hs}
\bibitem{Lautrup:1977hs}
B.~Lautrup,
%``On High Order Estimates In QED,''
Phys.\ Lett.\ {\bf B69} (1977) 109.
%%CITATION = PHLTA,B69,109;%%
%\cite{'tHooft:1977am}
%\bibitem{'tHooft:1977am}
%G.~'t Hooft,
%``The Whys of Subnuclear Physics'', 
%Lectures given at Int. School of Subnuclear Physics, Erice, Sicily, July 23 - 
%August 10, 1977.
%\cite{Zakharov:1992bx}
\bibitem{Zakharov:1992bx}
V.~I.~Zakharov,
%``QCD perturbative expansions in large orders,''
Nucl.\ Phys.\ {\bf B385} (1992) 452.
%%CITATION = NUPHA,B385,452;%%
%\cite{Brown:1992pk}
\bibitem{Brown:1992pk}
L.~S.~Brown, L.~G.~Yaffe and C.~Zhai,
%``Large order perturbation theory for the electromagnetic current-current correlation function,''
Phys.\ Rev.\ {\bf D 46} (1992) 4712
[hep-ph/9205213].
%%CITATION = HEP-PH 9205213;%%
%\cite{Krasnikov:1996jq}
\bibitem{Krasnikov:1996jq}
N.~V.~Krasnikov and A.~A.~Pivovarov,
%``Renormalization schemes and renormalons,''
Mod.\ Phys.\ Lett.\ {\bf A11} (1996) 835
[hep-ph/9602272].
%%CITATION = HEP-PH 9602272;%%
%\cite{Ilchev:1981mh}
\bibitem{Ilchev:1981mh}
A.~S.~Ilchev and V.~K.~Mitrjushkin,
%``On The Problem Of Summing The Perturbation Theory: A QED Example,''
J.\ Phys.\ G {\bf G7} (1981) L221.
%%CITATION = JPHGB,G7,L221;%%
%\cite{Broadhurst:1995se}
\bibitem{Broadhurst:1995se}
D.~J.~Broadhurst and A.~G.~Grozin,
%``Matching QCD and HQET heavy - light currents at two loops and beyond,''
Phys.\ Rev.\ {\bf D 52} (1995) 4082
[hep-ph/9410240].
%%CITATION = HEP-PH 9410240;%%
%\cit%\cite{Lovett-Turner:1995ti}
\bibitem{Lovett-Turner:1995ti}
C.~N.~Lovett-Turner and C.~J.~Maxwell,
%``All orders renormalon resummations for some QCD observables,''
Nucl.\ Phys.\ {\bf B452} (1995) 188
[hep-ph/9505224].
%%CITATION = HEP-PH 9505224;%%e{Lovett-Turner:1995ti}
%\cite{Gorishnii:1984cu}
\bibitem{Gorishnii:1984cu}
S.~G.~Gorishny, A.~L.~Kataev and S.~A.~Larin,
%``The Width Of Higgs Boson Decay Into Hadrons: Three Loop Corrections Of Strong Interactions,''
Sov.\ J.\ Nucl.\ Phys.\ {\bf 40} (1984) 329.
%%CITATION = SJNCA,40,329;%%
%\cite{Gorishnii:1990zu}
\bibitem{Gorishnii:1990zu}
S.~G.~Gorishny, A.~L.~Kataev, S.~A.~Larin and L.~R.~Surguladze,
%``Corrected Three Loop QCD Correction To The Correlator Of The Quark Scalar Currents And Gamma (Tot) (H0 $\to$ Hadrons),''
Mod.\ Phys.\ Lett.\ {\bf A5} (1990) 2703.
%%CITATION = MPLAE,A5,2703;%%
%\cite{Gorishnii:1991zr}
\bibitem{Gorishnii:1991zr}
S.~G.~Gorishny, A.~L.~Kataev, S.~A.~Larin and L.~R.~Surguladze,
%``Scheme dependence of the next to next-to-leading QCD corrections to Gamma(tot) (H0 $\to$ hadrons) and the spurious QCD infrared fixed point,''
Phys.\ Rev.\ {\bf D 43} (1991) 1633.
%%CITATION = PHRVA,D43,1633;%%
%\cite{Chetyrkin:1997sr}
\bibitem{Chetyrkin:1997sr}
K.~G.~Chetyrkin,
%``Correlator of the quark scalar currents and Gamma(tot)(H --> hadrons)  at O(alpha(s)**3) in pQCD,''
Phys.\ Lett.\ {\bf B390} (1997) 309
[hep-ph/9608318].
%%CITATION = HEP-PH 9608318;%%
%\cite{Broadhurst:1993si}
\bibitem{Broadhurst:1993si}
D.~J.~Broadhurst,
%``Large N expansion of QED: Asymptotic photon propagator and contributions to the muon anomaly, for any number of loops,''
Z.\ Phys.\ {\bf C58} (1993) 339.
%%CITATION = ZEPYA,C58,339;%%
%\cite{Shifman:1979bx}
\bibitem{Shifman:1979bx}
M.~A.~Shifman, A.~I.~Vainshtein and V.~I.~Zakharov,
%``QCD And Resonance Physics. Sum Rules,''
Nucl.\ Phys.\ {\bf B147} (1979) 385.
%%CITATION = NUPHA,B147,385;%%
%\cite{Surguladze:1990sp}
\bibitem{Surguladze:1990sp}
L.~R.~Surguladze and F.~V.~Tkachov,
%``Two Loop Effects In QCD Sum Rules For Light Mesons,''
Nucl.\ Phys.\ {\bf B331} (1990) 35.
%%CITATION = NUPHA,B331,35;%%
%\cite{Yndurain:1998gk}
\bibitem{Yndurain:1998gk}
F.~J.~Yndurain,
%``Pure QCD bounds and estimates for light quark masses,''
Nucl.\ Phys.\ {\bf B517} (1998) 324
[hep-ph/9708300].
%%CITATION = HEP-PH 9708300;%%
%\cite{Gabrielli:1993wv}
\bibitem{Gabrielli:1993wv}
E.~Gabrielli and P.~Nason,
%``Instanton effects in the light quark masses determination from QCD sum rules,''
Phys.\ Lett.\ {\bf B313} (1993) 430.
%%CITATION = PHLTA,B313,430;%%
%\cite{Chetyrkin:1998ej}
\bibitem{Chetyrkin:1998ej}
K.~G.~Chetyrkin, J.~H.~Kuhn and A.~A.~Pivovarov,
%``Determining the strange quark mass in Cabibbo suppressed tau lepton  decays,''
Nucl.\ Phys.\ {\bf B533} (1998) 473
[hep-ph/9805335].
%%CITATION = HEP-PH 9805335;%%
%\cite{Korner:2000wd}
\bibitem{Korner:2000wd}
J.~G.~Korner, F.~Krajewski and A.~A.~Pivovarov,
%``Determination of the strange quark mass from Cabibbo suppressed tau  decays with resummed perturbation theory in an effective scheme,''
hep-ph/0003165.
%%CITATION = HEP-PH 0003165;%%
%\cite{Grunberg:1984fw}
\bibitem{Grunberg:1984fw}
G.~Grunberg,
%``Renormalization Scheme Independent QCD And QED: The Method Of Effective Charges,''
Phys.\ Rev.\ {\bf D 29} (1984) 2315.
%%CITATION = PHRVA,D29,2315;%%
%\cite{Stevenson:1981vj}
\bibitem{Stevenson:1981vj}
P.~M.~Stevenson,
%``Optimized Perturbation Theory,''
Phys.\ Rev.\ {\bf D 23} (1981) 2916.
%%CITATION = PHRVA,D23,2916;%%
%\cite{Gardi:1998rf}
\bibitem{Gardi:1998rf}
E.~Gardi and M.~Karliner,
%``Relations between observables and the infrared fixed-point in {QCD},''
Nucl.\ Phys.\ B {\bf 529} (1998) 383
[hep-ph/9802218].
%%CITATION = HEP-PH 9802218;%%
%\cite{Mikhailov:1998xi}
\bibitem{Mikhailov:1998xi}
S.~V.~Mikhailov,
%``Renormalon chains contributions to non-singlet evolution kernels in  {QCD},''
Phys.\ Lett.\ B {\bf 431} (1998) 387
[hep-ph/9804263].
%%CITATION = HEP-PH 9804263;%%
%\cite{Mankiewicz:1997gz}
\bibitem{Mankiewicz:1997gz}
L.~Mankiewicz, M.~Maul and E.~Stein,
%``Perturbative part of the non-singlet structure function F2 in the large  N(F) limit,''
Phys.\ Lett.\ B {\bf 404} (1997) 345
[hep-ph/9703356].
%%CITATION = HEP-PH 9703356;%%
%\cite{Larin:1994vu}
\bibitem{Larin:1994vu}
S.~A.~Larin, T.~van Ritbergen and J.~A.~Vermaseren,
%``The Next next-to-leading QCD approximation for nonsinglet moments of deep inelastic structure functions,''
Nucl.\ Phys.\ B {\bf 427} (1994) 41.
%%CITATION = NUPHA,B427,41;%%






\end{thebibliography}
\end{document}